\documentclass{article}
\usepackage{cite}
\usepackage{graphicx}
\usepackage{dcolumn}

\begin{document}

\date{}
\title{Symmetry and degeneracy, exceptional point and coalescence: a pedagogical
approach}
\author{Francisco M. Fern\'{a}ndez\thanks{%
fernande@quimica.unlp.edu.ar} \\
INIFTA, DQT, Sucursal 4, C.C 16, \\
1900 La Plata, Argentina}
\maketitle

\begin{abstract}
We show a parameter-dependent $3\times 3$ non-Hermitian matrix that exhibits
both degeneracy and coalescence of eigenvalues at an exceptional point
(Hermitian and non-Hermitian degeneracies). This simple non-Hermitian model
is suitable for the discussion of those concepts in an undergraduate or
graduate course on quantum-mechanics. We also study the symmetry group
responsible for the degeneracy.
\end{abstract}

\section{Introduction}

\label{sec:intro}

Degeneracy is an important concept discussed in most textbooks on quantum
mechanics\cite{CDL77} and quantum chemistry\cite{P68} and several textbooks
on mathematics show its relationship with symmetry\cite{H62,C90}. In recent
years there has been great interest in non-Hermitian quantum mechanics\cite
{B05,B07} (and references therein) that gives rise to the concept of
exceptional points\cite{HS90,H00,HH01,H04,GRS07}, also known as defective
points\cite{MF80}, that also play a relevant role in perturbation theory\cite
{K95}. Non-Hermitian quantum mechanics and exceptional points have become so
relevant nowadays that there have been several pedagogical papers published
recently on the subject\cite{BBJ03,DV18,F18,LXLL20}.

The effect of exceptional points is most dramatically illustrated by
parameter-dependent Hamiltonians. As the model parameter approaches an
exceptional point two (or sometimes more) real eigenvalues approach each
other and coalesce. They emerge on the other side of the exceptional point
as a pair of complex-conjugate numbers. This coalescence is different from
degeneracy because at the exceptional point there is only one linearly
independent eigenvector. However, it is sometimes called non-Hermitian
degeneracy as opposed to Hermitian degeneracy\cite{BO98}.

The purpose of this paper is to illustrate the difference between
coalescence and degeneracy by means of a simple, exactly solvable
one-parameter model. In section~\ref{sec:Model} we discuss the model, in
section~\ref{sec:symmetry} we discuss degeneracy from the point of view of
symmetry and, finally, in section~\ref{sec:conclusions}~we summarize the
main results of the paper and draw conclusions.

\section{The model}

\label{sec:Model}

\begin{figure}[tbp]
\begin{center}
\includegraphics[width=9cm]{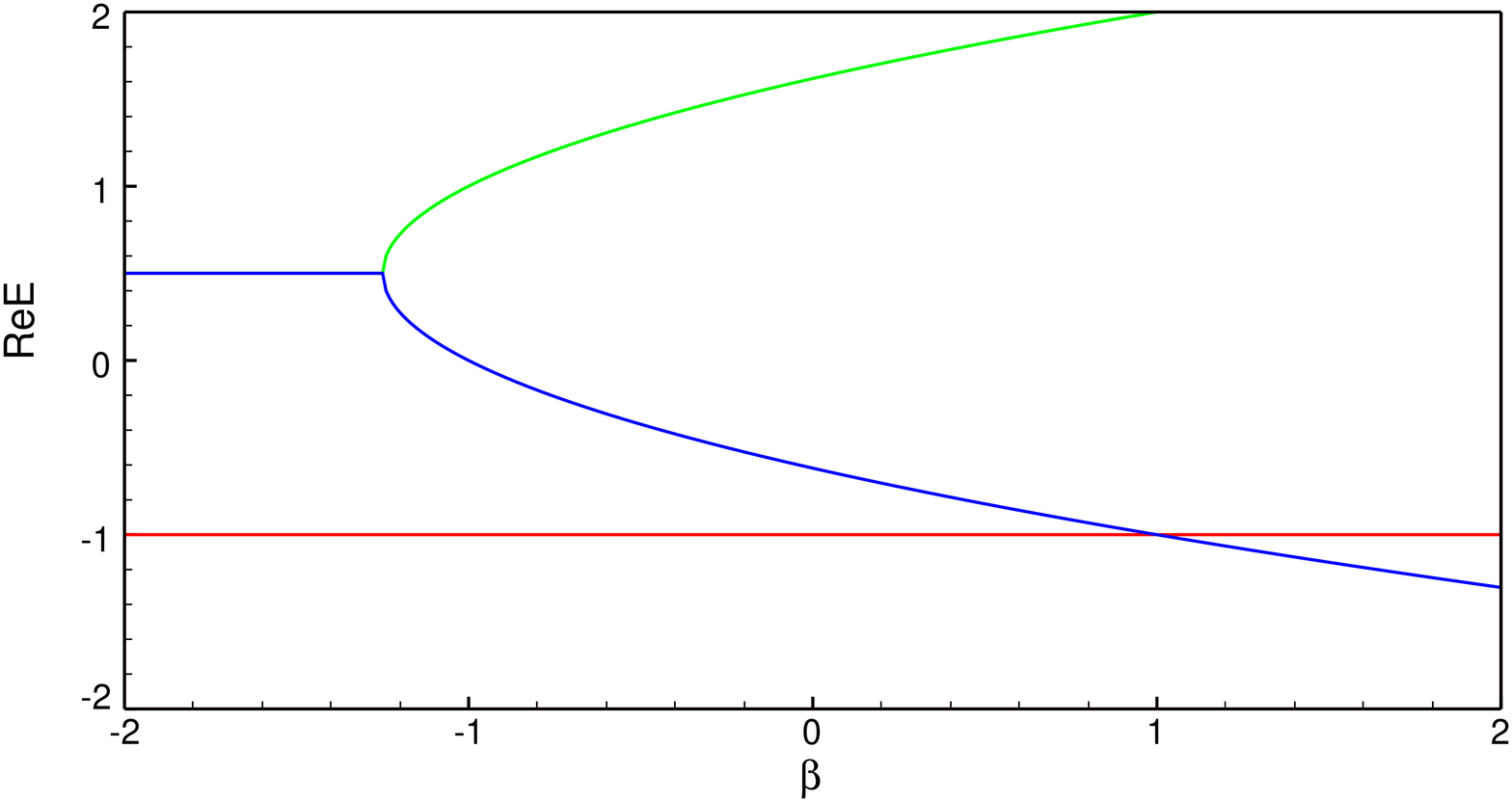}
\end{center}
\caption{Real part of the eigenvalus}
\label{Fig:ReE}
\end{figure}

\begin{figure}[tbp]
\begin{center}
\includegraphics[width=9cm]{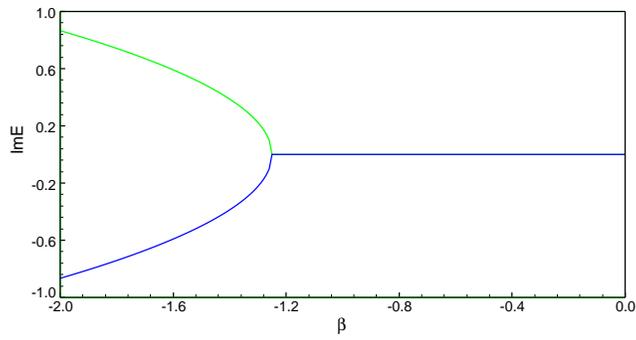}
\end{center}
\caption{Imaginary part of the eigenvalues}
\label{Fig:ImE}
\end{figure}

In order to illustrate both degeneracy and coalescence of eigenvalues we
propose the non-symmetric matrix
\begin{equation}
\mathbf{H}(\beta )=\left(
\begin{array}{lll}
0 & 1 & 1 \\
1 & 0 & 1 \\
\beta & 1 & 0
\end{array}
\right) ,  \label{eq:H}
\end{equation}
that has the following eigenvalues
\begin{eqnarray}
E_{1} &=&-1,\;E_{2}=\frac{1}{2}\left( 1-\sqrt{4\beta +5}\right) ,\;E_{3}=%
\frac{1}{2}\left( 1+\sqrt{4\beta +5}\right) ,\;\beta <1,  \nonumber \\
E_{1} &=&\frac{1}{2}\left( 1-\sqrt{4\beta +5}\right) ,\;E_{2}=-1,\;E_{3}=%
\frac{1}{2}\left( 1+\sqrt{4\beta +5}\right) ,\;\beta >1,
\label{eq:Eigenvalues(beta)}
\end{eqnarray}
labelled so that $E_{1}\leq E_{2}\leq E_{3}$. The real and imaginary parts
of these eigenvalues are shown in figures \ref{Fig:ReE} and \ref{Fig:ImE},
respectively.

We appreciate that $E_{1}$ and $E_{2}$ cross at $\beta =1$ and swap their
relative order. These eigenvalues become degenerate at $\beta =1$ and a set
of three orthonormal eigenvectors of $\mathbf{H}(1)$ are
\begin{eqnarray}
E_{1} &=&E_{2}=-1,\;\mathbf{v}_{1}=\frac{1}{\sqrt{6}}\left(
\begin{array}{c}
1 \\
1 \\
-2
\end{array}
\right) ,\;\mathbf{v}_{2}=\frac{1}{\sqrt{2}}\left(
\begin{array}{c}
1 \\
-1 \\
0
\end{array}
\right) ,  \nonumber \\
E_{3} &=&2,\;\mathbf{v}_{3}=\frac{1}{\sqrt{3}}\left(
\begin{array}{l}
1 \\
1 \\
1
\end{array}
\right) .  \label{eq:Eigenval_eigenvect_beta=1}
\end{eqnarray}
The symmetric matrix $\mathbf{H}(1)$ can be diagonalized by means of the
orthogonal matrix $\mathbf{C}$ constructed from the eigenvectors (\ref
{eq:Eigenval_eigenvect_beta=1}) in the usual way\cite{P68}:
\begin{equation}
\mathbf{C}^{t}\mathbf{H}(1)\mathbf{C=}\left(
\begin{array}{ccc}
-1 & 0 & 0 \\
0 & -1 & 0 \\
0 & 0 & 2
\end{array}
\right) ,\;\mathbf{C}=\frac{1}{6}\left(
\begin{array}{ccc}
\sqrt{6} & 3\sqrt{2} & 2\sqrt{3} \\
\sqrt{6} & -3\sqrt{2} & 2\sqrt{3} \\
-2\sqrt{6} & 0 & 2\sqrt{3}
\end{array}
\right) .  \label{eq:Diag_H(1)}
\end{equation}
Note that the matrix $\mathbf{H}(1)$ exhibits $3$ linearly independent
eigenvectors, two of them degenerate. Besides, $E_{1}$ and $E_{2}$ remain
real before and after the point $\beta =1$ as shown in figure \ref{Fig:ReE}.
This is the usual degeneracy commonly found in quantum mechanics\cite{CDL77}
and quantum chemistry\cite{P68}.

On the other hand, the eigenvalues $E_{2}$ and $E_{3}$ coalesce at $\beta
=-5/4$ and become a pair of complex conjugate numbers when $\beta <-5/4$
(see figures \ref{Fig:ReE} and \ref{Fig:ImE}). The matrix $\mathbf{H}(-5/4)$
exhibits eigenvalues and eigenvectors
\begin{eqnarray}
E_{1} &=&-1,\;\mathbf{v}_{1}=\frac{1}{\sqrt{2}}\left(
\begin{array}{c}
0 \\
1 \\
-1
\end{array}
\right) ,  \nonumber \\
E_{2} &=&E_{3}=\frac{1}{2},\;\mathbf{v}_{2}=\frac{1}{3}\left(
\begin{array}{c}
2 \\
2 \\
-1
\end{array}
\right) .  \label{eq:Eigenvalues_beta=-5/4}
\end{eqnarray}
In this case there are only two linearly independent eigenvectors
and the matrix $\mathbf{H}(-5/4)$ is defective
(non-diagonalizable). One can obtain other suitable vectors by
means of a Jordan chain\cite{GRS07} \newline(see,
https://en.wikipedia.org/wiki/Generalized\_eigenvector\#Jordan\_chains,
for
examples). In the present case we obtain a third column vector $\mathbf{v}%
_{3}$ with elements $c_{1}$, $c_{2}$ and $c_{3}$ from
\begin{equation}
\left[ \mathbf{H}\left( -\frac{5}{4}\right) -\frac{1}{2}\mathbf{I}\right]
\left(
\begin{array}{c}
c_{1} \\
c_{2} \\
c_{3}
\end{array}
\right) =\mathbf{v}_{2},  \label{eq:Jordan_chain}
\end{equation}
where $\mathbf{I}$ is the $3\times 3$ identity matrix. One possible solution
is
\begin{equation}
\mathbf{v}_{3}=\frac{1}{6}\left(
\begin{array}{c}
6 \\
6 \\
1
\end{array}
\right) .  \label{eq:Jordan_vector}
\end{equation}
With these three vectors we can convert $\mathbf{H}\left( -\frac{5}{4}%
\right) $ into a Jordan matrix
\begin{equation}
\mathbf{S}^{-1}\mathbf{H}\left( -\frac{5}{4}\right) \mathbf{S}=\left(
\begin{array}{c|cc}
-1 & 0 & 0 \\ \hline
0 & \frac{1}{2} & 1 \\
0 & 0 & \frac{1}{2}
\end{array}
\right) ,\;\mathbf{S}=\frac{1}{6}\left(
\begin{array}{ccc}
0 & 4 & 6 \\
3\sqrt{2} & 4 & 6 \\
-3\sqrt{2} & -2 & 1
\end{array}
\right) ,  \label{eq:Jordan_matrix}
\end{equation}
where the two Jordan blocks are explicitly indicated.

\section{Symmetry}

\label{sec:symmetry}

The matrix $\mathbf{H}(1)$ can be thought as a some kind of description of
three identical objects. Therefore, the six orthogonal matrices $\mathbf{U}%
_{i}$, $i=0,1,\ldots ,5$ that carry out the six permutations of three
objects $\left( c_{1}\;c_{2}\;c_{3}\right) $ should leave $\mathbf{H}(1)$
invariant. In order to construct such matrices we proceed as indicated in
what follows:
\begin{eqnarray}
\left(
\begin{array}{l}
c_{1}^{\prime } \\
c_{2}^{\prime } \\
\vdots \\
c_{N}^{\prime }
\end{array}
\right) &=&\mathbf{U}\left(
\begin{array}{l}
c_{1} \\
c_{2} \\
\vdots \\
c_{N}
\end{array}
\right) ,  \nonumber \\
c_{i}^{\prime } &=&\sum_{j=1}^{N}u_{ij}c_{j},\;u_{ij}=\frac{\partial
c_{i}^{\prime }}{\partial c_{j}},  \label{eq:U_construction}
\end{eqnarray}
where $u_{1j}$, $i,j=1,2,\ldots ,N$, are the matrix elements of $\mathbf{U}$%
. As an example, consider the cyclic permutation
\begin{eqnarray}
\left(
\begin{array}{l}
c_{3} \\
c_{1} \\
c_{2}
\end{array}
\right) &=&\mathbf{U}_{1}\left(
\begin{array}{l}
c_{1} \\
c_{2} \\
c_{3}
\end{array}
\right) ,  \nonumber \\
c_{1}^{\prime } &=&c_{3},\;c_{2}^{\prime }=c_{1},\;c_{3}^{\prime }=c_{2}.
\label{eq:U_construction_example}
\end{eqnarray}

In this way we construct the group of matrices $\left\{ \mathbf{U}_{0},\,%
\mathbf{U}_{1},\,\ldots ,\mathbf{U}_{5}\right\} $%
\begin{eqnarray}
\mathbf{U}_{0} &=&\left(
\begin{array}{ccc}
1 & 0 & 0 \\
0 & 1 & 0 \\
0 & 0 & 1
\end{array}
\right) ,\;\mathbf{U}_{1}=\left(
\begin{array}{ccc}
0 & 0 & 1 \\
1 & 0 & 0 \\
0 & 1 & 0
\end{array}
\right) ,\;\mathbf{U}_{2}=\left(
\begin{array}{ccc}
0 & 1 & 0 \\
0 & 0 & 1 \\
1 & 0 & 0
\end{array}
\right) ,  \nonumber \\
\mathbf{U}_{3} &=&\left(
\begin{array}{ccc}
1 & 0 & 0 \\
0 & 0 & 1 \\
0 & 1 & 0
\end{array}
\right) ,\;\mathbf{U}_{4}=\left(
\begin{array}{ccc}
0 & 0 & 1 \\
0 & 1 & 0 \\
1 & 0 & 0
\end{array}
\right) ,\;\mathbf{U}_{5}=\left(
\begin{array}{ccc}
0 & 1 & 0 \\
1 & 0 & 0 \\
0 & 0 & 1
\end{array}
\right) ,  \label{eq:group_matrices}
\end{eqnarray}
that leave the matrix $\mathbf{H}(1)$ invariant: $\mathbf{U}_{i}^{t}\mathbf{H%
}(1)\mathbf{U}_{i}=\mathbf{H}(1)$. The group of the six permutations of
three objects (including the identity $\mathbf{U}_{0}$) is commonly known as
the symmetric group $S_{3}$\cite{H62} that is isomorphic to $D_{3}$ and $%
C_{3v}$\cite{P68,C90}. When $\beta \neq 1$ the only matrix that leaves $%
\mathbf{H}(\beta )$ invariant is $\mathbf{U}_{0}$.

\section{Conclusions}

\label{sec:conclusions}

In this paper we compare two apparently similar concepts: degeneracy and
coalescence. Although such concepts have been discussed in the past, here we
propose a simple, exactly solvable model that exhibits both. In our opinion
this model is suitable for the discussion of these concepts in an
introductory course on quantum mechanics. In addition, this simple model is
also suitable for the illustration of the relationship between symmetry and
degeneracy. It is quite easy to construct the orthogonal matrices that
commute with the Hamiltonian one that becomes symmetric at $\beta =1$ and
exhibits the greatest degree of degeneracy. All the required algebraic
calculations can be more easily carried out by means of available computer
algebra software. For this reason this model is suitable for learning the
application of such tools.

\end{document}